\documentclass[10pt,preprint,longabstract]{aastex}

\def\msun{$M_{\odot}$}

\def\xte{J1550}

\begin{document}

\title{An Improved Dynamical Model for the Microquasar 
XTE J1550-564$^{\dagger}$}

\author{Jerome A. Orosz}
\affil{Department of Astronomy, San Diego State University,
5500 Campanile Drive, San Diego, CA 92182-1221}
\email{orosz@sciences.sdsu.edu}

\author{James F. Steiner, Jeffrey E. McClintock, Manuel A. P. Torres}
\affil{Harvard-Smithsonian Center for Astrophysics, 60 Garden Street,
Cambridge, MA 02138}
\email{jsteiner@cfa.harvard.edu,
jem@cfa.harvard.edu, mtorres@cfa.harvard.edu}

\author{Ronald A. Remillard}
\affil{Kavli Institute for Astrophysics and Space Research, 
Massachusetts Institute of Technology,
Cambridge, MA 02139-4307}
\email{rr@space.mit.edu}

\author{Charles D. Bailyn}
\affil{Department of Astronomy, Yale University, P.O. Box 208101,
New Haven, CT 06520}
\email{charles.bailyn@yale.edu}

\author{Jon M. Miller}
\affil{Department of Astronomy, University of Michigan,
500 Church Street, Dennison 814, Ann Arbor, MI 48109-1090}
\email{jonmm@umich.edu}

\altaffiltext{$\dagger$}{This 
paper includes data gathered with the 6.5 meter Magellan
Telescopes located at Las Campanas Observatory, Chile.}

\begin{abstract}
We present an improved dynamical model of the X-ray binary and
microquasar XTE J1550-564 based on new moderate-resolution optical
spectroscopy and near-infrared photometry obtained with the 6.5m
Magellan Telescopes at Las Campanas Observatory.  Twelve spectra of
the source were obtained using the Magellan Echellette Spectrograph
between 2008 May 6 and August 4.  In addition, several hundred images
of the field were obtained between 2006 May and 2009 July in the $J$
and $K_S$ filters using the PANIC camera.  The agreement between the
2006/2007 and 2008 $J$ and $K_S$ light curves is not perfect, and the
differences can plausibly be attributed to a hot spot on the accretion
disk during the 2006/2007 observations.  By combining our new radial
velocity measurements with previous measurements obtained 2001 May at
the 8.2m VLT and with light curves, we find an orbital period of
$P=1.5420333\pm 0.0000024$ days and a radial velocity semiamplitude of
$K_2=363.14\pm 5.97$ km s$^{-1}$, which together imply an optical mass
function of $f(M)=7.65\pm 0.38\,M_{\odot}$.  We find that the
projected rotational velocity of the secondary star is $55\pm 5$ km
s$^{-1}$, which implies a very extreme mass ratio of $Q\equiv
M/M_2\approx 30$.  Using a model of a Roche lobe-filling star and an
azimuthally symmetric accretion disk, we fit simultaneously optical
light curves from 2001, near-infrared light curves from 2008 and all
of the radial velocity measurements to derive system parameters.  We
find an inclination of $74.7\pm 3.8^{\circ}$ and component masses of
$M_2=0.30\pm 0.07\,M_{\odot}$ and $M=9.10\pm 0.61\,M_{\odot}$ for the
secondary star and black hole, respectively.  We note that these
results depend on the assumption that in 2008, the disk does not have a hot
spot, and that the fraction of light contributed by the accretion disk
did not change between the spectroscopic and photometric observations.
By considering two measured values of the disk fraction and by
modeling various combinations of NIR and optical light curves, we show
that our adopted black hole mass is probably not seriously in error,
where the black hole mass ranges between $M=8.91\pm 1.10\,M_{\odot}$
and $M=13.94\pm 1.64\,M_{\odot}$.  The radius of the secondary star
for the adopted model is $1.75\pm 0.12\,R_{\odot}$.  Using this
radius, the average $K_S$ magnitude, and an extinction of
$A_K=0.507\pm 0.050$ mag, we find a distance of $4.38^{+0.58}_{-0.41}$
kpc, which is in good agreement with a recent distance estimate based
on HI absorption lines.
\end{abstract}

\section{Introduction}

With the sole exception of Cygnus X-1, all of the 18 confirmed stellar
black holes in the Galaxy are transient X-ray sources \citep{rem06}.  
One of the most important and thoroughly-studied of
these transient X-ray binary systems is XTE J1550--564 (hereafter
J1550), which was discovered on 1998 September 6 \citep{smi98}.
Within a week, its intensity rose to $\sim1.5$ Crab, and on September
19--20, the source produced an impulsive and unprecedented flare that
reached a peak intensity of 6.8 Crab \citep{sob00}.
Following this remarkable $\lesssim2$-day event, the source returned
to its pre-flare intensity and remained at that level for a month
before fading to a state of near-obscurity.  Remarkably, in early
December the source again brightened.  Within a few weeks thereafter,
it reached an intensity of $\sim2.5$ Crab and maintained that level
for a few months before it slipped below the threshold of {\em RXTE} (a few
mCrab) in mid-1999 \citep{sob00}.  
A second and smaller
outburst was observed during 2000 April-June 
\citep{rod03},
and three subsequent, brief mini-outbursts were observed
in 2001, 2002, and 2003 \citep{rem06}.

The source J1550 is especially important because of the relativistic
phenomena it has exhibited.  It is one of seven sources of
high-frequency quasi-periodic oscillations (QPOs), and it is only one
of four such sources that exhibit harmonic pairs of frequencies in a
3:2 ratio (\citealt{mil01}; 
\citealt{re02a}; \citealt{rem06}).  
J1550 is also one of only three sources that has
been observed to produce a large-scale relativistic X-ray jet 
\citep{cor02}.  
This two-sided jet, observed in 2000 and 2002 by {\it  Chandra}, 
was presumably launched during the powerful 1998 X-ray
flare, which was observed to be accompanied by a superluminal radio
jet \citep{han09}.  
The rich X-ray timing phenomenology of
J1550 has also been important for studying the complex power spectra
of black hole transients, which are often punctuated by strong
low-frequency QPOs (\citealt{hom01}; \citealt{re02b}).

The optical counterpart of J1550 was identified shortly after the
discovery of the X-ray source \citep{oro98}
and extensive
photometric observations were made during both the 1998--1999 and 2000
outburst cycles \citep{jai99,ja01a,ja01b}.  
In May and June of
2001, in a state of deep quiescence, with the X-ray source dimmed by
more than a million-fold \citep{cor06}, 
we made optical
photometric and spectroscopic observations of the counterpart 
\citep{oro02}.  
The analysis of these data yielded a large mass
function, $f(M) = 6.86 \pm 0.71$~\msun, thereby clinching the argument
that the compact primary is a black hole.  The dynamical model we
developed is comprised of an $\approx 10$ \msun~black hole in a
1.54-day orbit with a late G or early K companion.

In this paper, we revisit our earlier model, bringing to bear
extensive near-infrared photometric data obtained during 2006--2009
and 12 echellete spectra obtained in 2008.  These spectra have a
resolving power $R=4200-4950$, more than twice the resolving power of
the spectra we used in our earlier analysis and modeling.  In 
\citet{oro02}, using the low-resolution spectra available to us then, we
adopted for the projected rotational velocity of the companion star
the ``tentative value of $V_{\rm rot}$sin$i=90\pm10$~km~s$^{-1}$.''
Now, using our higher-resolution spectra, we find $V_{\rm
  rot}$sin$i=55\pm5$~km~s$^{-1}$.  The effect of this new
determination of the rotational velocity is to very significantly
boost the mass ratio $Q \equiv M/M_2$ from $\sim 9$ to $\sim 30$.
This change accounts for most of the differences between the current
model and our earlier one.

In Section 2, we present our new near-infrared light curves and our
moderate resolution spectra.  In Section 3 we discuss the analysis of
the new data and the results we derive from them.  In Section 4 we
briefly discuss some of the implications for our results, and in
Section 5 we end with a brief summary.

\section{Observations and Reductions}

\subsection{Photometry}\label{photsec}

\xte\ was observed in the near-infrared (NIR) using the Persson's
Auxiliary Nasmyth Infrared Camera (PANIC; \citealt{mar04}) and
$J$, $H$, and $K_S$ filters on the Magellan Baade telescope at Las
Campanas Observatory.  These data were obtained over runs spanning
three years ranging from 2006 May to 2009 July.  The data were
primarily gathered in two observing sequences during 2007 May and 2008
April (see Table \ref{obstab} for details).  
To mitigate problems inherent in the NIR arising from hot or
dead pixels, the data were collected in sequences consisting of a
``dice-5'' dithering pattern with an offset between 6 and 8
arcseconds, with one or three exposures taken at each position.  Each
such group of observations have been combined for analysis using an
IRAF reduction package for the PANIC detector, after being
flat-fielded.  In order to stay within the linear range of the
detector, individual exposures were typically limited to 10s for all
three bands.  Dice-5 dithering sequences then are mostly either 50 or
150s in effective duration.  In addition, a few sequences with
exposure times of 3s were obtained in order to keep bright comparison
stars in the linear regime.

We used the PANIC reduction package to perform the reductions on the
images.  While reducing the PANIC data, it was necessary to skip the
default pipeline sky subtraction step. This step is designed to
produce a blank sky image using a pixel-by-pixel median along a
sequence of images. However, for crowded fields such as that of \xte,
this sky image actually can inadvertently contain stars and other
pattern noise. By omitting this step, the final images contained a
low-level, large-scale curvature in sky brightness over the
field. This effect had negligible impact on the psf-derived
photometry.

All images were corrected for spatial distortion and nonlinearity and rotated
and aligned to a common orientation.  The highest quality data were
stacked and combined to produce a deep field image from which a
catalogue of field stars was derived.  This list was used to uniformly
analyze all images with the {DAOPHOT} \citep{ste87}
package in IRAF.
Iterative selection produced approximately forty stars ranging from
1.0-4.5 mag brighter than \xte\ with no signs of variability.  These
stars were used to calibrate the field by employing the least-squares
differential photometric solution discussed in \citet{Honeycutt1992}.

Zero-point calibrations were derived using a set of 12 relatively
isolated field stars with available 2MASS \citep{skr06}
photometry, filtered to remove any clearly variable objects.  The
2MASS magnitudes were transformed to the LCO standard system following
\citet{Carpenter2001}.  These stars range from $\approx 12.5-14.5$ mag
in $K_S$ and $\approx 13-15$ mag in $J$.  A weighted average based
upon 2MASS uncertainties was used to determine the zero-point offsets,
with total uncertainty estimated from a quadrature sum of the weighted
error combined with a $\approx$ 1\% error describing the 2MASS
field-to-field piecewise zero-point
variability\footnote{http://www.ipac.caltech.edu/2mass/releases/allsky/doc/sec4\_8.html}.
The resultant uncertainties are 0.022, 0.019, and 0.019 mag in the
$J$, $H$, and $K_S$-bands, respectively.  The field near \xte\ is
relatively crowded (Figure 1).  In particular, there are two stars
close to \xte\ on the sky: A star dubbed ``A'' that is about 25\% as
bright as \xte\ and is located about 1 arcsecond to the northeast, and
a star dubbed ``B'' that is only 5-10\% as bright as \xte\ and is
about 0.7 arcseconds to the southeast.  Because DAOPHOT had a high
failure rate to solve for both \xte\ and B, B was omitted from the
DAOPHOT's input star list.  In order to assess any distortion in the
derived magnitudes due to the proximity of A and B, a set of 2500
simulated fields in $J$ and $K_S$ were produced with the average noise
and background level of a typical dice-5 sequence using an average
brightness of \xte.  Stars A and B, along with reference stars for
calibration, were also included in the simulation.  These simulations
were analyzed in the same manner as the data.  The scaling and
position of the stars were iterated and adjusted until close
($\lesssim 2$\%) agreement was achieved between the average counts
from a stack of 100 simulated images compared to a stack of 25 data
images at comparable and good seeing ($\lesssim 0\farcs5$).  The
simulations uniformly sampled a range of seeing from FWHM $\approx
0\farcs35 - 2\farcs0$, a broader range than that spanned by the data.
A step-like shift of $\sim 0.05$ mag appeared in the derived magnitude
when the seeing became comparable to the point-separation of \xte\ and
B (see Figure \ref{Jsimulation}).  These simulations were also used to
derive a seeing cut at $\approx 1\farcs2$, the confusion limit between
\xte\ and A.

This step-like impact on the 
\xte\ light curves by contamination from
B is well-described by a logistic curve of the form
\begin{equation}
\Delta=a+{b\over 1+\exp[-(x-c)/d]},
\end{equation}
where the variable $x$ is the seeing FWHM.  This was fit to the
simulations, achieving a goodness-of-fit of $\chi^2/\nu \approx 1.2$
for both $J$ and $K_S$-bands.  Seeing-based corrections for the data
have been derived from these models and applied, scaling by the
inverse of the observed flux (relative to band average).  The
corrections are typically 1\%-2\% and 5\% at maximum.  The uncertainty
from this correction is small ($\sim 1\%$) and has been estimated and
added in quadrature to the measurement errors.  The $H$-band data were
too sparse to derive a comparable phase-averaged correction. However,
the noise and background level for $H$ was very close to $K_S$, and so
the $K_S$ model was also used to correct the $H$-band data.  (Note
that the difference between $J$ and $K_S$ corrections is $\lesssim
1\%$ over the range of useful data, so any systematic effect from the
use of a $K_S$-band model will be very small.)

For the final adopted light curves, data were screened to fulfill
three conditions: (1) the seeing FWHM had to be $<1\farcs2$, which led
to the rejection of 5\% of the images.  To minimize confusion with A,
we further required (2) that the magnitude of A fall within 4 times
the median absolute deviation (MAD) of its average magnitude in each
band.  Failure to meet this criterion eliminated an additional
$\approx 5$\% of data in $J$ and $\approx 7$\% in $K_S$.  Lastly, (3)
data were only acccepted if the scatter of the reference star
calibrations for a given image was less than $4\sigma$ above the
average for each band.  In the end, approximately 85\% of the $J$- and
$K$-band images (187 out of 223 for $J$ and 362 out of 422 for $K_S$)
and all 17 $H$-band images were selected.

The $J$ and $K_S$ light curves are shown in Figure \ref{plotallpanic}
phased on the ephemeris derived below.  The ellipsoidal modulation is
evident in the light curves and shows up quite clearly in the $J$-band
light curves from 2008.  The agreement between the light curves from
2006/2007 and the light curves from 2008 is generally good, although
the incomplete phase coverage hinders this comparison.  The light
curves disagree somewhat near orbital phase 0.0, which corresponds to
the inferior conjunction of the companion star.  The light curves near
this phase appear to be ``filled in'' during the 2006-2007 period.
Apart from the difference near phase zero, there does not seem to be
any significant change in the mean brightness level from one year to
the next.  The modest differences that do exist are likely
attributable to changes in the disk.  A similar but much greater
degree of disk variability in quiescence has been seen in the black
hole candidate A0620-00 \citep{Cantrell2009}.

\subsection{Spectroscopy}\label{specsec}

We obtained moderate resolution spectra of \xte\ using the Magellan
Echellette Spectrograph (MagE; Marshall et al.\ 2008) and the Clay
telescope at LCO on the nights of 2008 May 6, June 28, and August 3-4.
For the May observations the slit and resolving power were
respectively $1\farcs0$ and $R=4200$; for all other observations the
slit was $0\farcs85$ and $R=4950$.  The skies were clear and the
seeing was usually $\le 0\farcs8$.  The exposure times were typically
1200 seconds each, and ThAr lamp exposures were obtained frequently
throughout the observing runs.  The images were reduced with tasks in
the IRAF `ccdproc' and `echelle' packages.  After the bias was
subtracted from each image, pairs of images were combined using a
clipping algorithm to remove cosmic rays.  This process yielded three
spectra from May, three spectra from June, and 13 spectra from August.
The resulting images were flat-fielded using a normalized master flat
and then rotated in order to align the background night sky emission
lines along the columns.  After the background emission lines were
rectified, the spectra from individual orders were optimally
extracted.  The signal-to-noise ratio of the extracted spectra were
typically between 5 and 10 per pixel in the best echelle order.

\section{Analysis and Results}\label{resultsec}

\subsection{Radial Velocities of the Secondary Star}\label{radvel}

For the MagE spectra of \xte\ there is a single echelle order that
covers the region between the interstellar sodium D lines and
H$\alpha$ that we used to determine radial velocities.  The bluer
orders deliver poor signal-to-noise because the source is quite red,
and the redder orders are unusable because of heavy fringing and
strong telluric features.  The `fxcor' routine in IRAF (based on the
cross-correlation technique given in \citealt{ton1979}) was used to
measure the radial velocities.  We used a spectrum of the K3III star
HD 181110 as the template.  The same template star was used by 
\citet{oro02}
for the analysis of their 8.2m VLT spectra.  The three
MagE spectra from May gave good cross-correlation peaks, as did nine
of the spectra from June and August.  The values of the Tonry \& Davis
'$r$' parameter were generally between 3 and 5 for these spectra.  In
the analysis below we used only the 12 spectra from which usable
radial velocity measurements were derived.

It proved to be difficult to combine the radial velocities obtained by
the VLT (17 observations) and by Magellan (12 observations).  The
velocities from each data set were fitted to a sine curve of the form
$V(\phi)=K\sin(\phi)+\gamma$, where $\phi$ is the orbital phase.  Even
though the same template star was used for both analyses, the
$\gamma$-velocities differed by 70 km s$^{-1}$.  By using common stars
that were observed by both telescopes, it was discovered that the
zero-point of the VLT velocities is uncertain by at least 30 km
s$^{-1}$.  In addition, it is possible that the template star is a
single-lined binary star with a moderately long period, resulting in a
small change in its radial velocity between the two runs.  To combine
the two data sets, we simply removed the fitted $\gamma$-velocity
appropriate for each set from each individual radial velocity.  For
the combined set containing 29 radial velocities, a 
sine curve was fitted yielding an orbital period of $P = 1.5420435 \pm
0.0000069$ days and a radial velocity semiamplitude of $K_2 = 365.5
\pm 7.1$ km s$^{-1}$.  The radial velocities and the best-fitting sine
curve are shown in Figure \ref{rvfig1}.  In our earlier work 
\citep{oro02}, 
the VLT spectra alone yielded $K_2 = 349 \pm 12$ km
s$^{-1}$, and fitting both the light-curve and velocity data gave
$352.2\le K_2 \le 370.1$ km s$^{-1}$ ($1\sigma$). 
Although the VLT observations
were obtained shortly after a weak minioutburst, the source appeared to
be very near X-ray quiescence 
\citep{oro02}.  This impression seems to be confirmed when the Magellan
and radial velocities are compared in Figure \ref{rvfig1}, where 
there appears to be no systematic differences
between the Magellan and VLT radial velocities.  

\subsection{The Rotational Velocity of the Secondary Star
and the Disk Fraction}\label{rotsec}

In order to measure the projected rotational velocity of the secondary
star and the fraction of light contributed by the accretion disk (the
``disk fraction''), we made a ``restframe'' spectrum from the nine
spectra obtained in June and August (i.e.\ spectra that show
cross-correlation peaks and were also obtained with the $0\farcs85$
slit).  Each spectrum (from the single echelle order) was normalized
to its continuum using a polynomial fit and Doppler shifted to zero
velocity.  The resulting nine normalized spectra were then averaged
with no clipping or rejection of bad pixels.  The ``broadening
function'' technique developed by \citet{ruc02}
was used to measure
the projected rotational velocity of the secondary.  The spectrum of
the K3III star was used as the high signal-to-noise sharp-lined
reference spectrum.  The broadening kernel\footnote{Basically, the
  broadening kernel is the function needed to convert a hypothetical
  spectrum of a non-rotating star into its observed restframe
  spectrum.} (hereafter the Broadening Function or BF) appropriate for
the restframe spectrum was derived, and is shown in Figure
\ref{BFfig}.  The BF from the rest frame spectrum is well-fitted by a
Gaussian with a FWHM of $65.4\pm 2.5$ km s$^{-1}$.

We then performed a few simple numerical experiments in order to map
the relationship between the BF FWHM and the rotational velocity of
the star.  A spectrum of the K4 III star HD 181480 (observed using
MagE and the $0\farcs85$ slit) was normalized to its continuum and
rotationally broadened by various rotational velocities using the
familiar analytic broadening kernel (e.g.\ \citealt{gray92})
with a limb darkening coefficient of $\epsilon=0.6$.  The BFs of the
rotationally broadened spectra were derived using the same template
that was used for the \xte\ restframe spectrum.  The plot of the input
rotational velocity vs.\ the Gaussian FWHM of the derived BFs is shown
in Figure \ref{calibfig}.  For large rotational velocities, the
Gaussian FWHM roughly gives the input rotational velocity, as
expected.  Not surprisingly, the curve begins to flatten out at low
velocities where the input rotational velocity is comparable to the
spectral resolution.  The observed Gaussian FWHM for \xte\ of $65.4\pm
2.5$ km s$^{-1}$ seems to imply a rather low rotational velocity of
around 50 km s$^{-1}$.  In our previous work \citep{oro02}, a
provisional value of about 90 km s$^{-1}$ was derived from spectra of
lower resolution (3.6~\AA\ FWHM).  The MagE results clearly do not
support this.   
The width of the BFs from the individual spectra are
narrow (to the extent that one can measure them).  One expects that
the process of making a restframe spectrum would result in a BF that
is as wide or wider than the individual BFs.  Thus it seems very
unlikely that the narrow BF width is an artifact of the process that
was used to make the restframe spectrum.  Since the actual rotational 
velocity of the companion star
presumably has not changed, it appears that 
the value of 90 km s$^{-1}$ derived from the VLT was spurious and due to 
the low resolving power of the averaged spectrum.

Next, we used the technique outlined in \citet{mar94}
to decompose the restframe spectrum into its stellar and disk
components.  This was the same technique we used in our previous paper
\citep{oro02}.  Briefly, a template spectrum is normalized to
its continuum, and the Doppler shift between it and the
\xte\ restframe spectrum is removed.  The template spectrum is scaled
by a weight $w$ and subtracted from the \xte\ restframe spectrum.  A
polynomial is fit to the residuals and the rms of the fit is recorded.
The value of $w$ that gives the lowest rms is taken to be the optimal
weight for that template.  The four template stars that gave the best
decompositions for the VLT spectra \citep{oro02}
were
re-observed with MagE and the $0\farcs85$ slit.  The spectral types of
the templates are G8IV, K1III, K3III, and K4III.  The K3III template
gave the lowest rms using $w=0.61$, and the G8IV template gave the
worse rms values.  Based on this, we adopt a spectral type of K3III,
with an uncertainty of perhaps 1 subclass.

The K3III spectrum was then rotationally broadened by various amounts
between 40 and 90 km s$^{-1}$, and the decomposition process was repeated.
For each input value of $V_{\rm rot}\sin i$ the minimum rms was recorded.
Figure \ref{vrotfig} shows the rms vs.\ the input rotational velocity.
The rms is at its minimum value at an input rotational velocity of
$V_{\rm rot}\sin i=57$ km s$^{-1}$.  As was the case with the BF analysis,
a rotational velocity as high as 90 km s$^{-1}$ is clearly ruled out.

Based on the BF analysis and on the results shown in Figure
\ref{vrotfig}, we adopt a $1\sigma$ range of the projected rotational
velocity of $50\le V_{\rm rot}\sin i\le 60$ km s$^{-1}$.  Assuming the
secondary star is rotating synchronously with its orbit and that it
fills its Roche lobe, there is a relatively straightforward relation
between the observed (projected) rotational velocity of the star, its
$K$-velocity, and the mass ratio of the binary (e.g.\ \citealt{wad88}).  
We show from the dynamical modeling discussed below that the
projected rotational velocity of the star cannot be much below 50 km
s$^{-1}$, assuming a reasonable mass for the star.  Our adopted values
of $V_{\rm rot}\sin i=55\pm 5$ km s$^{-1}$ and $K_2=365.5\pm 7.1$ km
s$^{-1}$ imply a rather extreme mass ratio of $Q\equiv M/M_2\approx
30$.

Figure \ref{dffig} shows the rms vs.\ $w$ for the K3III template,
rotationally broadened using $V_{\rm rot} \sin i=57$ km s$^{-1}$.  The
minimum of the curve is well-defined, and gives a {\em disk} fraction
of 0.39, where the disk fraction is taken to be $1-w$.  The BF of the
smoothest residual spectrum from the decomposition was computed and is
shown in Figure \ref{BFfig}.  There is no significant peak in the BF,
which is a good indication that the stellar absorption lines have been
removed.

We adopt a disk fraction of $0.39\pm 0.05$, at an effective wavelength
of 6200~\AA, which we take to be the $R$-band in the ellipsoidal
modeling discussed below.  For comparison, \citet{oro02}
adopted a $V$-band disk fraction of $k_V=0.30 \pm 0.05$.  We believe
the $R$-band value of $k_R=0.39\pm 0.05$ is more reliable than the
$V$-band determination of $k_V=0.30 \pm 0.05$ owing to the better
resolution of the MagE spectra.

\subsection{Temperature of the Secondary Star and Extinction}\label{tempsec}

The spectral type of the star can be used to estimate its
temperature. However, the temperature corresponding to a given
spectral type depends on the luminosity class, with main-sequence
stars having higher temperatures than the corresponding giants.  For
example, \citet{str81}
give $T_{\rm eff}\approx
4700$ K for a K3V star, and $T_{\rm eff}\approx 4250$ K for a K3III
star.  A variation of $\pm1$ spectral subtype implies a temperature in
the range $T_{\rm eff}\gtrsim4100$ K (the K4III star) and $T_{\rm
  eff}\lesssim 4850$ K (the K2V star).  However, the surface gravity
of the secondary star in \xte\ is larger than the nominal surface
gravity of a ``normal'' K3III star.  \citet{str81}
give $\log g\approx 2.38$ for a K3III star.  The surface gravity of
the star in \xte\ is well determined from the dynamical model
discussed below and is $\log g=3.43$.  For the nominal main sequence
K3V star \citet{str81}
give $\log g\approx 4.56$.
If we do a simple interpolation in $\log g$, the tables of
\citet{str81}
would give $T_{\rm eff}\approx 4475$
K for a K3 star with a surface gravity matching the \xte\ secondary.

The NIR is especially useful for determining the intrinsic color of
the star and the reddening to the source because it only very weakly
depends on the assumed reddening law (see \citealt{dra03}).  Thus,
using NIR data one can place simultaneous constraints upon the
extinction to the source and the spectral type of the star.
\citet{Vuong2003} has directly measured the relationship between X-ray
absorption and NIR extinction using several young stellar objects
embedded in the nearby molecular cloud $\rho$ Ophiuchi.  For each
source, they measure $N_{H}$ from X-ray edge absorption and determine
$J$-band extinction $A_{J}$.  They find
\begin{equation}
A_{J}(\rm mag)={N_{H}\over (5.6\pm0.4) \times 10^{21}{\rm cm}^{-2}},
\end{equation}
which is significantly below the optically determined Galactic
relationship (e.g.\ rescaling the fit of
\citet{pre1995}
to NIR bands gives
$N_{H}/A_{J}=6.75\times10^{21}$cm$^{-2}$mag$^{-1}$).  Vuong
and collaborators find that this apparent discrepancy is completely
explainable by nonstandard abundances present in the local ISM (using
values taken from \citealt{wil00}).

Since the NIR reddening law depends only modestly on the reddening
parameter $R_{V}$, NIR colors are relatively insensitive to the
line-of-sight dust properties.  We obtain model extinction estimates
using the full NIR extinction curves modeled by the ``fmcurve''
routines of \citet{Fitzpatrick1999} and the filter properties of
PANIC \citep{mar04}.  We assume average Galactic reddening
$R_{V}=3.1$ \citep{car1989} to convert between
$A_{V}$ and $A_{J,H,K}$ and note that the differences
introduced into $A_{J}/A_{K}$ by the full range of $R_{V}$
is a negligible source of error.

We used a Monte Carlo approach to determine extinction, while
incorporating all relevant uncertainties.  Adopting $N_{H} =
(8.0\pm0.4)\times 10^{21}{\rm cm}^{-2}$ (90\% confidence), as measured
by \citet{Miller2003}, and assuming a Gaussian systematic uncertainty
on $N_{H}/A_{J}$ of $\sigma_{\rm sys} =
0.6\times10^{21}$cm$^{-2}$mag$^{-1}$ (i.e., one-half the difference
arising from using local vs.\ standard Galactic ISM abundances), we
obtained $A_{J}=1.17\pm0.11$, $A_{H} =0.762 \pm 0.074 $, and
$A_{K}=0.507\pm0.050$.

The observed NIR color of the secondary star is $J-K_S=1.27\pm 0.04$.
The unreddened color would then be $(J-K_S)_0=0.610\pm 0.083$ using
$A_K=0.507\pm 0.050$.  We have used zero-point uncertainties of 0.02
mag for both $J$ and $K_S$ to find the uncertainty in $(J-K_S)_0$.  In
order to find the temperature of the secondary star from the observed
$(J-K_S)_0$, we used synthetic photometry from the {\sc NextGen}
models \citep{hau99a,hau99b}.  
Using these models, we can
compute the $J-K_S$ color of the secondary star given its temperature
and gravity. 
Owing to the nature of Roche geometry, the gravity is strongly constrained
by the ellipsoidal models (discussed below)
for a wide range of temperatures, and we adopt
$\log g=3.43$ for these calculations.  
We found
from extensive experimentation with the synthetic ellipsoidal light
curves discussed below that the addition of an accretion disk that
satisfies the constraint of a disk fraction of $0.39$ at
6200~\AA\ does not significantly alter the $J-K_S$ color of the star,
{\em despite the fact that the computed disk fractions in $J$ and $K$
  are not zero}.  The top of Figure \ref{plottempvscolor} shows the
value of $(J-K_S)_0$ as a function of temperature.  The observed value
of $(J-K_S)_0=0.610$ implies a temperature of about 4450 K, in good
agreement with the temperature derived from the spectral type above.
The $1\sigma$ range on the color maps into a temperature range of
$4300\lesssim T_{\rm eff}\lesssim 4650$ K.

We did a Monte Carlo simulation to derive the distribution of allowed
temperatures, given the observed $J-K_S$ color of $1.27\pm 0.04$.  We
varied the model input temperature in steps of one degree between 3800
and 5900 K.  At each input temperature, 40,000 synthetic reddened
$J-K_S$ colors were constructed by drawing values of $A_K$ and
zero-point uncertainties from Gaussian distributions of the
appropriate width.  This process left us with about 84,000,000 $T_{\rm
  eff}, J-K_S$ pairs.  We then selected the pairs with a color between
$1.265\le J-K_S\le 1.275$ and computed a frequency distribution of the
temperatures from that interval.  The resulting histogram of
temperatures is shown in the bottom of Figure \ref{plottempvscolor}.
The distribution has two main peaks near 4450 and 5000K, with the most
likely temperature near 4450 K.  The interval of $4207\le T_{\rm
  eff}\le 5068$~K contains 68\% of the probability with equal
probabilities at each end.  Thus based on the color alone,
temperatures around 5100 K or so cannot be ruled out with high
confidence.

As noted above, the spectral type of the star implies a temperature
near 4475 K for a K3III spectral type and about 4700 for a K3V
spectral type.  These temperatures are near the leftmost peak in the
histogram in Figure \ref{plottempvscolor}.  For temperatures greater
than about 5200 K, the spectral type would be G8 or earlier for
luminosity class V, which can be ruled out based on the observed
spectral type.  For the purposes of the ellipsoidal modeling discussed
next, we adopt a temperature range of $4200 \le T_{\rm eff} \le 5200$
K, where the lower end is roughly the $1\sigma$ lower limit determined
from Figure \ref{plottempvscolor} and the upper end is imposed by the
spectral type.

\subsection{Simultaneous Fits to Light and Velocity Curves}\label{ELC}

We used the ELC code of \citet{oro2000}
to derive a
dynamical model of \xte.  The reader is referred to \citet{oro02}
for a thorough description of the assumptions and
techniques used.  The data modeled include optical light curves ($V$,
$g^{\prime}$, $r^{\prime}$, $i^{\prime}$, and $z^{\prime}$ from 
\citealt{oro02}), 
NIR light curves ($J$ and $K_s$ discussed here), and
radial velocities from the VLT \citep{oro02} 
and Magellan.  We
have three extra constraints: the disk fraction, the rotational
velocity of the star, and the fact that the X-ray source is not
eclipsed.  We used ELC's genetic and Monte Carlo Markov Chain
optimizers to arrive at the solutions.  We have 10 free parameters in
the model (the parameter ranges searched are given in parentheses):
the inclination $i$ (50.0--81.0 deg), the mass of the donor star $M_2$
(0.1--$3.0\,M_{\odot}$), the $K$-velocity of the donor star $K_2$
(336--396 km s$^{-1}$), the effective temperature of the donor star
$T_2$ (4200-5200 K), the orbital period $P$ (1.5420-1.5421 days), the
time of the inferior conjunction of the donor star [HJD 2,452,000 +
  (53.8--54.2)], and four parameters to specify the accretion disk,
namely the radius scaled to the Roche lobe radius (0.60--1.0), the
opening angle of the rim (0.5--12.0 deg), the temperature of the inner
rim (1500--20000 K), and the exponent on the radial temperature
profile (-0.9--0.0).  We have found that the search of parameter space
is more efficient if the mass of the donor star and its $K$-velocity
are used to set the scale of the binary instead of using the mass
ratio $Q$ and the orbital separation $a$.  We assume a circular orbit
and synchronous rotation.  We also assume that X-ray heating can be
neglected in the quiescent state (discussed further below).

Because of differences between the $J$ and $K_S$ light curves between
2006/2007 and 2008 (Figure \ref{plotallpanic}), and because we have
two measurements of the disk fraction (the $V$-band measurement in
\citealt{oro02}
and the $R$-band measurement discussed here), we
fit the data using eight combinations: optical light curves only,
optical light curves and the 2006/2007 NIR light curves, optical light
curves and the 2008 NIR light curves, and optical light curves plus
all NIR light curves.  For each of these four cases, we had two
constraints on the disk fraction: $k_V=0.30 \pm 0.05$ 
\citep{oro02}
and $k_R=0.39\pm 0.05$ derived here.  For the sake of the
discussion we label these eight cases as models A, B, ... H (see Table
\ref{parm}).  For each combination the genetic code was run three times from
different starting locations, and the Markov code was run twice.  For
each case a total of 250,000 to nearly one million models were
computed.   When run for a sufficient number of iterations, 
both the genetic code
and the Markov chain are insensitive to the precise ranges for the parameters, 
provided
the actual best-fitting parameters are within the specified ranges.
The ranges used for the fitted parameters were chosen to be large enough
so that they contained the model with the lowest $\chi^2$.
Owing to the multiple runs of the optimizers and the large
number of models computed, we are confident that the global minimum in
$\chi^2$ was reached in each case.

The results of the fitting are shown in Table \ref{parm}.  There are
some features in the table that should be noted.  First, the
$K$-velocities found from each situation are slightly different
(although well within their respective $1\sigma$ errors).  This is
because the computation of the radial velocity curve in ELC is not
completely decoupled from the light curves (e.g.\ the model atmosphere
intensities are tabulated in terms of the gravity $\log g$ which is
given in physical units), and because ELC also computes corrections to
the radial velocity curves caused by the distortion of the secondary
star (\citealt{wil76}; \citealt{eat08}).  Second, the range in the
inclination values is not terribly large.  The lowest inclination is
$57.7\pm 4.3^{\circ}$ for model C (i.e.\ the combination of the
2006/2007 NIR light curves and the optical light curves with
$k_V=0.30\pm 0.05$) and the largest inclination is $77.1\pm
7.1^{\circ}$ for model B (the combination of the optical light curves
with $k_R=0.39\pm 0.05$).  Hence the range of derived black hole
masses is relatively constrained: $(8.9\pm 1.1\,M_{\odot}) < M <
(13.9\pm 1.6\,M_{\odot})$.  Third, since the same optical light curves
are fitted in all 8 cases, the $\chi^2$ values for the fits to the
optical light curves give us a way to judge the relative
goodness-of-fit of each situation.  The situation that gives the best
fits to the optical light curves is for model B (optical curves only
and $k_R=0.39\pm 0.05$).  When we consider the situations where NIR
light curves are also fit, model F has the best fits to the optical
light curves.  Fitting the 2006/2007 NIR light curves with the optical
light curves (models C and D) or the combined NIR light curves and the
optical light curves (models G and H) results in significantly worse
fits to the optical data.  Thus it seems that the shapes of the
optical light curves are more consistent with the shapes of the 2008
NIR light curves.  Considering the cases where the 2008 NIR light
curves are fit with the optical light curves (models E and F), the
range in the inclination (and hence the black hole mass) are further
constrained: $(66.4\pm 6.9^{\circ}) < i < (74.7\pm 3.8^{\circ})$ and
$(9.1\pm 0.6\,M_{\odot}) < M < (10.6\pm 1.3\,M_{\odot})$.

As noted above, we believe our measurement of the $R$-band disk
fraction of $0.39 \pm 0.05$ is more secure than the $V$-band disk
fraction adopted by \citet{oro02} 
owing to the higher
resolution of the Magellan spectra.  In most cases, using the $R$-band
disk fraction as an extra constraint gives better fits to the optical
data compared to when the $V$-band disk fraction constraint is used.
Looking at cases when the $R$-band disk fraction constraint was used
(models D, F, and H), and where NIR data was used, model F gives the
best overall fit to the optical data.  Thus, for the final values of
the fitted and derived parameters, we adopt the results derived from
model F.  Figure \ref{fitfig} shows the phased light curves and the
best-fitting models.

Finally, we note that we have assumed the light curves have symmetry
in phase about $\phi=0.5$, and that the overall level of light has
stayed more or less constant from year to year.  Both these
assumptions may not be strictly true.  \citet{Cantrell2008} have
shown that A0620-00 has different optical ``states'' while in X-ray
quiescence, ranging from the relatively quite ``passive'' state to the
``active'' state where the brightness level and activity levels are
elevated relative to the passive state.  In addition, \citet{Cantrell2009}
have shown that even in the passive state the optical
light and NIR curves of A0620-00 can change with time.  These authors
showed that a bright spot on the accretion disk can account for most
of the observed changes.  It is difficult to directly assess the
importance of asymmetry or X-ray heating in the light curves,
especially in the NIR.  On the other hand, as noted above, the spread
in the inclination angles for all models is not that large, so we do
not expect the masses we report to have unreasonably large systematic
errors.

\subsection{Distance}

We compute the distance to the source using the synthetic photometry
code based on the {\sc NextGen} models.  Given a stellar radius,
gravity, and temperature, we compute the absolute magnitude of the
star in the $K_S$ filter.  We adjust the absolute magnitude to account
for the light from the accretion disk in the $K_S$ band.  Dereddening
the apparent $K_S$ magnitude, the distance then follows from the
difference between the absolute and apparent magnitudes.

The distance and its uncertainty for each model in Table 
\ref{parm}
was
computed using a simple Monte Carlo code.  For all models, we use an
apparent magnitude of $K=16.15\pm 0.04$, where we have used a generous
uncertainty to account for the zero-point error and the error in
determining the mean magnitude from the ellipsoidal curve, an
extinction of $A_K=0.507\pm 0.050$ (Section \ref{tempsec}), and a
stellar temperature drawn from the distribution shown in Figure
\ref{plottempvscolor}.  For individual models, the stellar radius, the
gravity, and the computed $K$-band disk fraction and their
uncertainties are listed in Table 
\ref{parm}.  Models A and B are fits to the
optical data only, and as such do not have computed $K$-band disk
fractions.  For our adopted model F, the top panel of Figure
\ref{distfig} shows the derived probability distribution of the
distance in kpc.  The most likely distance is 4.38 kpc, with a
$1\sigma$ range of $3.97 \le d \le 4.96$.  For comparison 
\citet{oro02}
derived a distance of $3.9\pm 1.8$ kpc for a secondary
star mass of $0.5\,M_{\odot}$.  Our revised distance is much more
precise because our dynamical model is more precise and because we are
using $K_S$ band observations where the extinction is minimal.

As an independent check, we computed the distance for the parameters
from model F using the apparent $J$-band magnitude ($J=17.42\pm 0.04$)
and extinction ($A_J=1.17=\pm 0.11$).  The bottom panel of Figure
\ref{distfig} shows the probability distribution for the distance in
kpc.  The most likely distance is 4.52 kpc, which is very close to the
most likely distance derived using the $K_S$-band.  However, owing to
the larger uncertainty in the $J$-band extinction, the distance
derived from $J$-band measurements has a larger uncertainty.  The
$1\sigma$ range for the $J$-band measurement is $3.87\le d\le 5.36$
kpc, which is nearly as large as the 90\% range on the $K_S$-band
measurement.

Based on a reanalysis of the radio images obtained during the 1998
outburst, \citet{han09} constrained the source distance
to be between 3.3 and 4.9 kpc based on absorption features seen in the
radio spectrum.  Our $1\sigma$ distance range is compatible with their
range.  Using $d=4.4$ kpc, the apparent separation velocity between
two ejection events observed in 1998 is $1.7c$, and the intrinsic
velocity is $\gtrsim 0.9c$ \citep{han09}.  
\citet{cor02} 
observed the radio jets in June of 2000 and again in
January of 2002.  They measured an average proper motion for the
approaching jet of $32.9\pm 0.7$ milliarcseconds (mas) day$^{-1}$ and
an average proper motion for the receding jet of $18.3\pm 0.7$ mas
day$^{-1}$.  At a distance of 4.4 kpc, these proper motions correspond
to large apparent velocities of $0.84c$ and $0.47c$.  \citet{cor02} 
also detected material from the approaching jet at X-ray
wavelengths in 2000 and in 2002.  The proper motions of the X-ray jet
were $21.2 \pm 7.2$ mas day$^{-1}$ between 2000 June and September,
and $10.4\pm 0.9$ mas day$^{-1}$ between 2000 September 2002 and
March.  At a distance of 4.4 kpc, these proper motions correspond to
apparent separation velocities of $0.55c$ and $0.27c$, respectively.
Our best-fit values also imply that the remarkable X-ray flare of 
1998 September 19-20 (see \S 1) reached a luminosity of $0.4 L_{\rm Edd}$,
as determined from dead-time corrected spectral analyses
of the {\it RXTE} pointed observation near MJD 51076.016.

\section{Discussion}\label{discuss}

\subsection{The Mass of the Black Hole and the Binary Mass Ratio}

The adopted value for the mass of the black hole is $9.10\pm
0.61\,M_{\odot}$ (Table \ref{parm}).  For comparison, \cite{oro02}
derived a mass of $9.56\pm 1.2\,M_{\odot}$ without using a constraint
on the (projected) rotational velocity of the star, and $10.63\pm
0.95\,M_{\odot}$ when using a tentative value of $V_{\rm rot}\sin
i=90\pm 10$ km s$^{-1}$ as an extra constraint.  Using the rotational
velocity as an extra constraint restricts the range of allowed mass
ratios, and $V_{\rm rot}\sin i=90$ km s$^{-1}$ gives $Q\approx 9$.  We
have shown here that the tentative value of $V_{\rm rot}\sin i=90\pm
10$ km s$^{-1}$ suggested by \citet{oro02}
is ruled out by the
moderate resolution Magellan spectra.  We have measured a much lower
and definite value of $V_{\rm rot}\sin i=55\pm 5$ km s$^{-1}$, which
implies a much more extreme mass ratio.  For our adopted model F, $Q
=30.1\pm 5.7$.  The mass ratio of \xte\ is one of the most extreme
measured for a black hole binary. Only XTE J1118+480 has a more
extreme value, with $Q\approx 42$ \citep{cav09}.

Both XTE J1550-564 and XTE J1118-480 showed periodic modulations
known as superhumps
in their light curves as they approached quiescence (Jain et al.\
2001a; Zurita et al.\ 2006).  Superhumps are thought to be caused
by the precession of a tidally distorted disk in a system
with an extreme mass ratio (Whitehurst 1988;
Whitehurst \& King 1991), and as noted above, 
both systems have very large mass ratios.  
Zurita et al.\ (1996) observed a 0.02 mag modulation 
during the early decline towards
quiescence, which they attributed to a superhump.  They also
noted slight distortions in quiescent light curves obtained on two
different nights, and these distortions were attributed  
to a residual superhump modulation.  

Could the quiescent light curves of XTE J1550-564 also have some kind
of residual superhump modulation?   While it is true that the
quiescent light curves of XTE J1118+480 shown in Zurita et al.\
(2006) had night-to-night changes, these changes may in fact be due to
other reasons.  For example, 
Cantrell et al.\
(2010) showed that quiescent light curves of A0620-00 can 
systematically change from night to night owing to changes in the
accretion disk (mainly the location of the hot spot).  As we noted in
Section 3.4 above, it is hard to completely rule out night-to-night
variations in the quiescent XTE J1550-564 light curves, whatever their cause.
Given that we have light curves from different seasons, we don't expect
our dynamical model to have large systematic errors.

Our refined measurement of the black hole mass, orbital inclination
and the distance to \xte\ are important for the measurement of the
black hole spin and for the interpretation of the observed
high-frequency X-ray QPOs.  A full discussion of the spin of this
source as derived from detailed X-ray spectral modeling will appear in
a subsequent paper (Steiner et al., in preparation).  We will briefly
discuss here the implications for the QPOs, which have been observed
in \xte\ at frequencies up to 284 Hz.  For a Schwarzschild black hole,
the frequency of the innermost stable circular orbit (ISCO) would be
at $\nu=2199/(M_{\rm BH}/M_{\odot})=242$ Hz for $M_{\rm BH}=9.10
\,M_{\odot}$ \citep{st83}.  
\citet{str01}
pointed
out that one of the high-frequency QPOS for GRO J1655-40 seemed to
imply substantial spin for the black hole, since its frequency would
be higher than the dynamical frequency of the accretion disk at the
ISCO for the case of zero spin.  By analogy, we might expect this to
occur as well for \xte, given that the high-frequency QPOs appear to
be similar, while both systems are transient sources of relativistic
jets (\citealt{han09}; \citealt{cor02}; \citealt{hje95}). 
However, this is not the case for \xte, since our new
mass determination allows a dynamical frequency at the ISCO (for the
case of zero spin) which is faster than the 276 Hz QPO seen from this
source \citep{re02a}.  
That is, $\nu=279$ when
$M=7.88\,M_{\odot}$, which is $2\sigma$ smaller than its nominal best
value.

\subsection{Evolutionary Status of the Secondary Star}\label{evol}

The mass donor star in \xte\ must be highly evolved owing to its
relatively low mass of $M_2=0.30\pm 0.07\,M_{\odot}$ and large radius
of $R_2=1.75\pm 0.12\,R_{\odot}$ (see 
Table \ref{parm}).  We show here that the
star follows the relatively simple ``stripped giant'' evolution in a
similar manner to the donor star in V404 Cyg \citep{king93}.  
In this
picture, the properties of the star depend very strongly on the core
mass $M_c$ and not on its total mass $M_2$.  \citet{web83}
show that the mass and radius of the star depend on
$M_c$ as
\begin{eqnarray}
\ln(L_2/L_{\odot})&=&3.50
                  + 8.11\ln(M_c/0.25)
                  - 0.61[\ln(M_c/0.25)]^2
                   -2.13[\ln(M_c/0.25)]^3 \\
\ln(R_2/R_{\odot})&=&2.53
                  + 5.10\ln(M_c/0.25)
                  - 0.05[\ln(M_c/0.25)]^2
                   -1.71[\ln(M_c/0.25)]^2, 
\end{eqnarray}
where $M_c$ is in units of $M_{\odot}$ and is in the range
$0.17\lesssim M_c\lesssim 0.45$.  A core mass of $M_c=0.1661$ is
needed to get a radius of $1.75\,R_{\odot}$, although we note that
this core mass is just below the nominal range over which the fitting
relations apply.  The luminosity at this core mass is
$0.946\,L_{\odot}$, which is consistent with the value we derived
($1.05^{+0.60}_{-0.33}\,L_{\odot}$).  Using Eqn.\ (33) of \citet{king88},
the mass transfer rate is
\begin{equation}
-\dot{M}_2\approx 5.4\times 10^{-9}\left(M_c\over 0.25\right)^{7.11}
(M_2/M_{\odot}) \approx 9\times 10^{-11}\,M_{\odot}
{\rm yr}^{-1}.\label{mdot}
\end{equation}
\citet{oro02}
attempted to estimate the mass transfer rate from
X-ray spectral models.  They derived a total X-ray fluence in the
2-100 keV band of $\approx 1.1$ ergs cm$^{-2}$ during 1998--2001.
Since that time, \xte\ has had minor outbursts in January of 2002
(\citealt{swa02}; \citealt{bel02}) and March of
2003 (\citealt{dub03}; 
\citealt{kul03}; 
\citealt{are04}; 
\citealt{stu05}).  As {\rm RXTE} All-Sky Monitor
data show, these events were much weaker in the soft X-rays (2-12 keV)
than the events between 1998 and 2001 \citep{rem06}.
For a distance of 4.4 kpc, the isotropic energy release during
1998--2001 was about $2.52\times 10^{45}$ ergs.  If we assume a
radiative efficiency of 10\% and also assume that the mass lost from
the secondary star is all captured by the black hole, then a
recurrence time between the major outburst episodes of about 157 years
would be needed for the mass transfer rate given in Equation
\ref{mdot}.  This recurrence time would increase if we considered the
additional energy release in the 2002 and 2003 events.  If the black
hole has a significant spin, then the efficiency of accretion can be
higher, which would lead to a smaller recurrence time, all other
things being equal.

As the core evolves, it grows in mass.  According to \citet{king88}, 
the
final core mass will be $M_c^f=M_2^f=(M_c^0)^{0.75}\approx
0.26\,M_{\odot}$.  For a present-day mass of $M_2=0.3\,M_{\odot}$ and
a core mass of $M_c=0.1661$, \xte\ can continue to accrete at its
present rate (Equation \ref{mdot}) for another $\approx 4.4\times
10^8$ years.  After the mass transfer has stopped, the final orbital
period of the binary will be about 44 days.

\section{Summary}\label{summary}

Using new moderate-resolution optical spectroscopy and near-infrared
photometry obtained with the 6.5m Magellan Telescopes, we have derived
a much improved dynamical model of the X-ray binary and microquasar
\xte.  By combining our new radial velocity measurements with the 17
measurements obtained 2001 May at the 8.2m VLT and with light curves,
we found an orbital period of $P=1.5420333\pm 0.0000024$ days and a
radial velocity semiamplitude of $K_2=363.14\pm 5.97$ km s$^{-1}$,
which together imply an optical mass function of $f(M)=7.65\pm
0.38\,M_{\odot}$.  We find that the projected rotational velocity of
the secondary star is much smaller than the tentative value found by
\citet{oro02}.  
Our new value is $V_{\rm rot}\sin i=55\pm 5$ km
s${_1}$, which implies a very extreme mass ratio of $Q\equiv
M/M_2\approx 30$.  We modeled simultaneously the optical and
near-infrared light curves and the radial velocity curve to derive
system parameters.  We found component masses of $M_2=0.30\pm
0.07\,M_{\odot}$ and $M=9.10\pm 0.61\,M_{\odot}$ for the secondary
star and black hole respectively.  The radius of the secondary star is
$1.75\pm 0.12\,R_{\odot}$.  Using this radius, the average $K_S$
magnitude, and an extinction of $A_K=0.507\pm 0.050$ mag, a distance
of $4.38^{+0.58}_{-0.41}$ kpc is derived.

\acknowledgments

We thank Danny Steeghs for the use of telescope
time at LCO.  JFS thanks Paul Martini for his gracious support with
the PANIC pipeline.  This publication makes use of data products from
the Two Micron All Sky Survey, which is a joint project of the
University of Massachusetts and the Infrared Processing and Analysis
Center/California Institute of Technology, funded by the National
Aeronautics and Space Administration and the National Science
Foundation.  The work of
JAO was supported in part by the National
Science Foundation grant
AST-0808145.
CBD gratefully acknowledges support from the National
Science Foundation grant AST-0707627.  The work of JEM was
supported in part by NASA grant NNX08AJ55G.  JFS was supported by
Smithsonian Institution Endowment Funds.  RR acknowledges support from
NASA via the contract for the instrument team of {\em RXTE} at MIT.

\clearpage

\begin{deluxetable}{lllll}
\tablecaption{Summary of Observations\label{obstab}}
\tablewidth{0pt}
\tablehead{
\colhead{UT Date} &
\colhead{Telescope} &
\colhead{Instrument} &
\colhead{Filter or Resolution} &
\colhead{Note}}
\startdata
2001 May 24-27  & VLT Antu 8.2m           & FORS1 & $V$  & Orosz et al.\ 2002 \\
2001 May 24-27  & VLT Antu 8.2m           & FORS1 & 3.6~\AA\ FWHM & Orosz et al.\ 2002 \\
2001 June 1     & ESO NTT  3.5m           & SuSI2  &  $V$, $R$           & Orosz et al.\ 2002 \\
2001 June 26-28 & Magellan Baade 6.5m & MagIC & $r^{\prime}$, $i^{\prime}$, $z^{\prime}$& Orosz et al.\ 2002  \\
2006 May 6      & Magellan Baade 6.5m & PANIC   & $K_s$     \\
2007 May 23-25  & Magellan Baade 6.5m & PANIC   & $J$, $H$, $K_s$     \\
2007 June 2-4, 21 & Magellan Baade 6.5m & PANIC   & $J$, $K_s$     \\
2008 April 14-16 & Magellan Baade 6.5m & PANIC   & $J$, $H$, $K_s$     \\
2008 May 6 & Magellan Clay 6.5m & MagE & 1.41~\AA\ FWHM & \\
2008 June 28 & Magellan Clay 6.5m & MagE & 1.21~\AA\ FWHM & \\
2008 August 3-4 & Magellan Clay 6.5m & MagE & 1.21~\AA\ FWHM & \\
2009 July 15 & Magellan Baade 6.5m & PANIC   & $K_s$     \\
\enddata
\end{deluxetable}

\thispagestyle{empty} 
\begin{deluxetable}{rccccccccc}
\tabletypesize{\scriptsize}
\rotate
\tablecaption{\xte\ Adopted Parameters\label{parm}}
\tablewidth{0pt}
\tablehead{
\colhead{parameter} &          
\multicolumn{2}{c}{optical only}  &   
\multicolumn{2}{c}{2006/7}  &   
\multicolumn{2}{c}{2008}     &
\multicolumn{2}{c}{combined}  &   
\colhead{adopted}             \\
\colhead{  }                  &
\colhead{$k_V=0.30$}              &
\colhead{$k_R=0.39$}                &
\colhead{$k_V=0.30$}              &
\colhead{$k_R=0.39$}                &
\colhead{$k_V=0.30$}            &
\colhead{$k_R=0.39$}              &
\colhead{$k_V=0.30$}                &
\colhead{$k_R=0.39$}            &
\colhead{value}     \\
\colhead{model designation}                  &
\colhead{A}              &
\colhead{B}                &
\colhead{C}              &
\colhead{D}                &
\colhead{E}            &
\colhead{F}              &
\colhead{G}                &
\colhead{H}            &
\colhead{ }     
}
\startdata
$P$ (days)             
                       &  $1.5420341$
                       &  $1.5420341$
                       &  $1.5420396$
                       &  $1.5420394$
                       &  $1.5420335$
                       &  $1.5420333$
                       &  $1.5420366$
                       &  $1.5420369$ 
                       &  $1.5420333$  \\ 
             
                       &  ~$\pm 0.0000036$
                       &  ~$\pm 0.0000031$
                       &  ~$\pm 0.0000031$
                       &  ~$\pm 0.0000055$
                       &  ~$\pm 0.0000037$
                       &  ~$\pm 0.0000024$
                       &  ~$\pm 0.0000027$
                       &  ~$\pm 0.0000051$ 
                       &  ~$\pm 0.0000024$  \\ 
$T_0$ (HJD 2,450,000+)\tablenotemark{a} 
                       &  $2053.9334$
                       &  $2053.9333$
                       &  $2053.9313$
                       &  $2053.9318$  
                       &  $2053.9301$
                       &  $2053.9306$  
                       &  $2053.9301$
                       &  $2053.9297$  
                       &  $2053.9306$ \\ 

                       &  ~$\pm 0.0036$
                       &  ~$\pm 0.0038$
                       &  ~$\pm 0.0061$
                       &  ~$\pm 0.0041$  
                       &  ~$\pm 0.0037$
                       &  ~$\pm 0.0039$  
                       &  ~$\pm 0.0030$
                       &  ~$\pm 0.0045$  
                       &  ~$\pm 0.0039$ \\ 
$K_2$ (km s$^{-1}$)    
                       & $364.55\pm 7.23$
                       & $364.33\pm 6.20$
                       & $366.55\pm 8.62$
                       & $368.18\pm 8.08$  
                       & $362.90\pm 6.22$
                       & $363.14\pm 5.97$  
                       & $364.91\pm 6.85$
                       & $365.35\pm 6.07$  
                       & $363.14\pm 5.97$   \\ 
$i$ (deg)              
                       & $63.39\pm 4.76$
                       & $77.09\pm 7.06$
                       & $57.72\pm 4.30$ 
                       & $65.50\pm 4.75$ 
                       & $66.37\pm 6.90$ 
                       & $74.69\pm 3.79$ 
                       & $66.25\pm 3.99$ 
                       & $70.97\pm 3.73$ 
                       & $74.69\pm 3.79$    \\
$M_2$ ($M_{\odot}$)    
                       & $0.39\pm 0.09$
                       & $0.30\pm 0.08$
                       & $0.48\pm 0.11$ 
                       & $0.36\pm 0.12$ 
                       & $0.37\pm 0.08$ 
                       & $0.30\pm 0.07$ 
                       & $0.41\pm 0.05$ 
                       & $0.31\pm 0.10$ 
                       & $0.30\pm 0.07$        \\
$R_2$ ($R_{\odot}$)    
                       & $1.90\pm 0.13$
                       & $1.74\pm 0.17$
                       & $2.04\pm 0.16$ 
                       & $1.85\pm 0.18$  
                       & $1.87\pm 0.12$ 
                       & $1.75\pm 0.12$  
                       & $1.92\pm 0.12$ 
                       & $1.76\pm 0.17$  
                       & $1.75\pm 0.12$           \\ 
$\log g$ (cgs)         
                       & $3.471\pm 0.035$
                       & $3.432\pm 0.049$
                       & $3.501\pm 0.040$ 
                       & $3.458\pm 0.045$ 
                       & $3.465\pm 0.037$ 
                       & $3.434\pm 0.033$ 
                       & $3.478\pm 0.040$ 
                       & $3.435\pm 0.046$ 
                       & $3.434\pm 0.033$   \\
$a$ ($R_{\odot}$)      
                       & $12.84\pm 0.57$
                       & $11.77\pm 0.46$
                       & $13.67\pm 0.56$ 
                       & $12.72\pm 0.53$ 
                       & $12.49\pm 0.54$ 
                       & $11.85\pm 0.28$ 
                       & $12.60\pm 0.63$ 
                       & $12.14\pm 0.31$ 
                       & $11.85\pm 0.28$  \\
$M$ ($M_{\odot}$)      
                       & $11.57\pm 1.48$
                       & ~$8.91\pm 1.10$
                       & $13.94\pm 1.64$ 
                       & $11.27\pm 1.47$ 
                       & $10.64\pm 1.32$ 
                       & ~$9.10\pm 0.61$ 
                       & $10.89\pm 1.68$ 
                       & ~$9.81\pm 0.74$ 
                       & ~$9.10\pm 0.61$   \\
$Q$ ($M/M_2$)      
                       & $29.6\pm 6.6$
                       & $29.8\pm 7.7$
                       & $29.0\pm 8.0$ 
                       & $31.5\pm 5.3$ 
                       & $28.5\pm 6.9$ 
                       & $30.1\pm 5.7$ 
                       & $26.9\pm 4.6$ 
                       & $32.0\pm 8.8$ 
                       & $30.1\pm 5.7$   \\
luminosity ($L_{\odot}$)
                       & $1.23^{+0.71}_{-0.39}$
                       & $1.02^{+0.65}_{-0.34}$
                       & $1.38^{+0.88}_{-0.41}$
                       & $1.16^{+0.70}_{-0.39}$
                       & $1.19^{+0.72}_{-0.36}$
                       & $1.05^{+0.60}_{-0.33}$
                       & $1.25^{+0.74}_{-0.38}$
                       & $0.99^{+0.68}_{-0.30}$
                       & $1.05^{+0.60}_{-0.33}$ \\
disk fraction ($J$)      
                       & \nodata
                       & \nodata
                       & $0.229\pm 0.051$ 
                       & $0.338\pm 0.048$ 
                       & $0.196\pm 0.052$ 
                       & $0.267\pm 0.034$ 
                       & $0.283\pm 0.045$ 
                       & $0.348\pm 0.026$ 
                       & $0.267\pm 0.034$  \\
disk fraction ($K$)      
                       & \nodata
                       & \nodata
                       & $0.243\pm 0.058$ 
                       & $0.336\pm 0.075$ 
                       & $0.176\pm 0.086$ 
                       & $0.236\pm 0.038$ 
                       & $0.303\pm 0.054$ 
                       & $0.362\pm 0.032$ 
                       & $0.236\pm 0.038$  \\
computed distance (kpc)
                       & \nodata
                       & \nodata
                       & $5.24^{+0.68}_{-0.55}$
                       & $5.07^{+0.70}_{-0.63}$
                       & $4.53^{+0.57}_{-0.41}$
                       & $4.38^{+0.58}_{-0.41}$
                       & $5.07^{+0.58}_{-0.41}$
                       & $4.85^{+0.71}_{-0.55}$
                       &  $4.38^{+0.58}_{-0.41}$ \\
$\chi^2$ 
    (optical)\tablenotemark{b}  
                       & 141.64
                       & 139.69 
                       & 155.67
                       & 157.58    
                       & 147.48    
                       & 146.81    
                       & 153.37    
                       & 154.58    
                       & \nodata         \\
$\chi^2$ 
    ($J$)\tablenotemark{c}  
                       & \nodata
                       & \nodata
                       & 105.78
                       & 104.29    
                       & 89.57    
                       & 86.74    
                       & 190.23    
                       & 189.25    
                       & \nodata         \\
$\chi^2$ 
    ($K$)\tablenotemark{c}  
                       & \nodata
                       & \nodata
                       & 278.21
                       & 279.32    
                       & 81.92   
                       & 82.02    
                       & 364.89    
                       & 363.23    
                       & \nodata         \\
$\chi^2$ 
 (radial velocities)   
                       & 26.04
                       & 26.19
                       & 28.32          
                       & 28.29    
                       & 25.98           
                       & 25.92    
                       & 25.48           
                       & 25.32  
                       & \nodata             \\
\enddata
\tablecomments{The effective temperature of the secondary has
been constrained to the range $4200\le T_{\rm eff}\le 5200$~K.}
\tablenotetext{a}{The time of inferior conjunction of the secondary star.}
\tablenotetext{b}{The sum of the $\chi^2$ values for the $V$, $g^{\prime}$,
$r^{\prime}$, $i^{\prime}$, and $z^{\prime}$ light curves.}  
\tablenotetext{c}{The number of $J$-band data points is $N_J=102,85,187$ for 2006/7, 2008, and combined,
respectively.  The number of $K$-band data points is $N_K=278,84,362$ 
for 2006/7, 2008, and combined,
respectively.}
\end{deluxetable}

\clearpage

\begin{figure}
\includegraphics[scale=0.7,angle=-90]{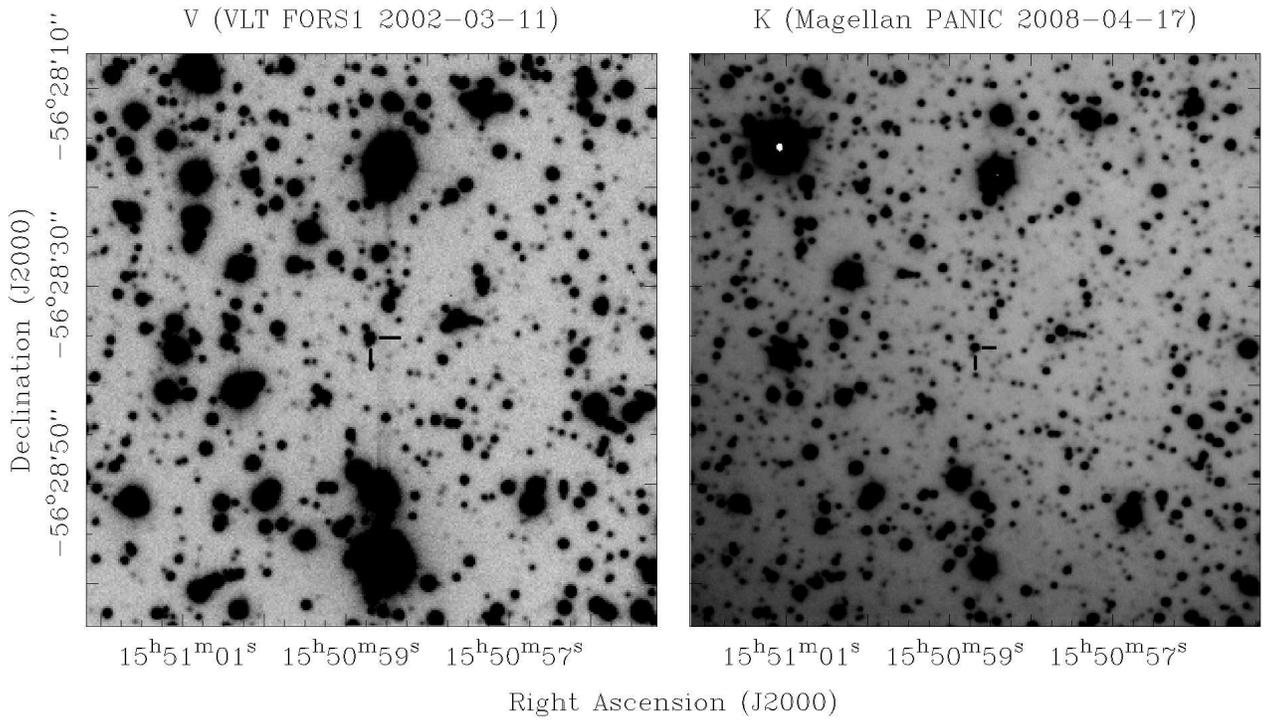}
\caption{$V$-band (left) and $K$-band (right) finding charts of the
field of \xte.  Each image is approximately $1^{\prime}\times
1^{\prime}$.  The $V$ band image is a 300 second exposure obtained
2002 March 11 with the VLT/FORS1 instrument in 0\farcs6 seeing.  The
composite $K$-band image was obtained with the Magellan/PANIC in
seeing conditions typically $\le 0\farcs6$.  The net exposure time
is 3000 seconds.}
\label{findingchart}
\end{figure}

\clearpage

\begin{figure}
\epsscale{0.9}
\plotone{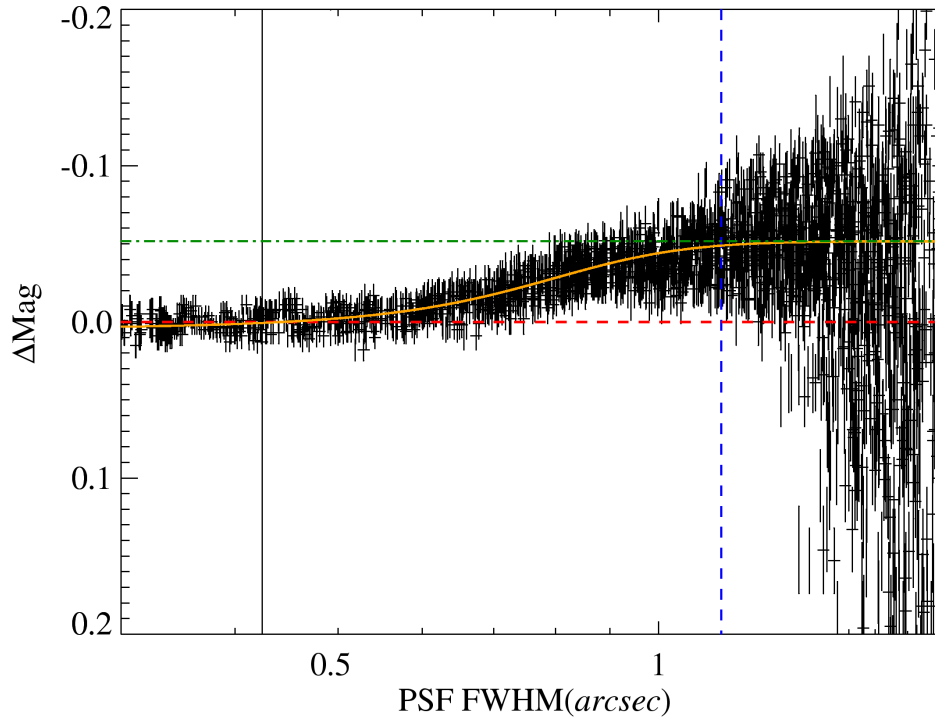}
\caption{Simulated offsets in the $J$-band magnitude of
\xte\ vs.\ seeing derived from the Monte Carlo simulations.  There
is a prominent dependence of the derived magnitude on seeing, with
an impact of up to ~0.05 mag (denoted by the green dot-dashed line).
The solid orange line, which is a fit to the trend, was used to
correct the instrumental magnitudes of \xte\ based on the seeing of
the image.  The vertical dashed blue line denotes the seeing cut
adopted for the data, and the vertical black line denotes the best
seeing during all of the runs.}
\label{Jsimulation}
\end{figure}

\clearpage

\begin{figure}
\includegraphics[scale=0.7,angle=-90]{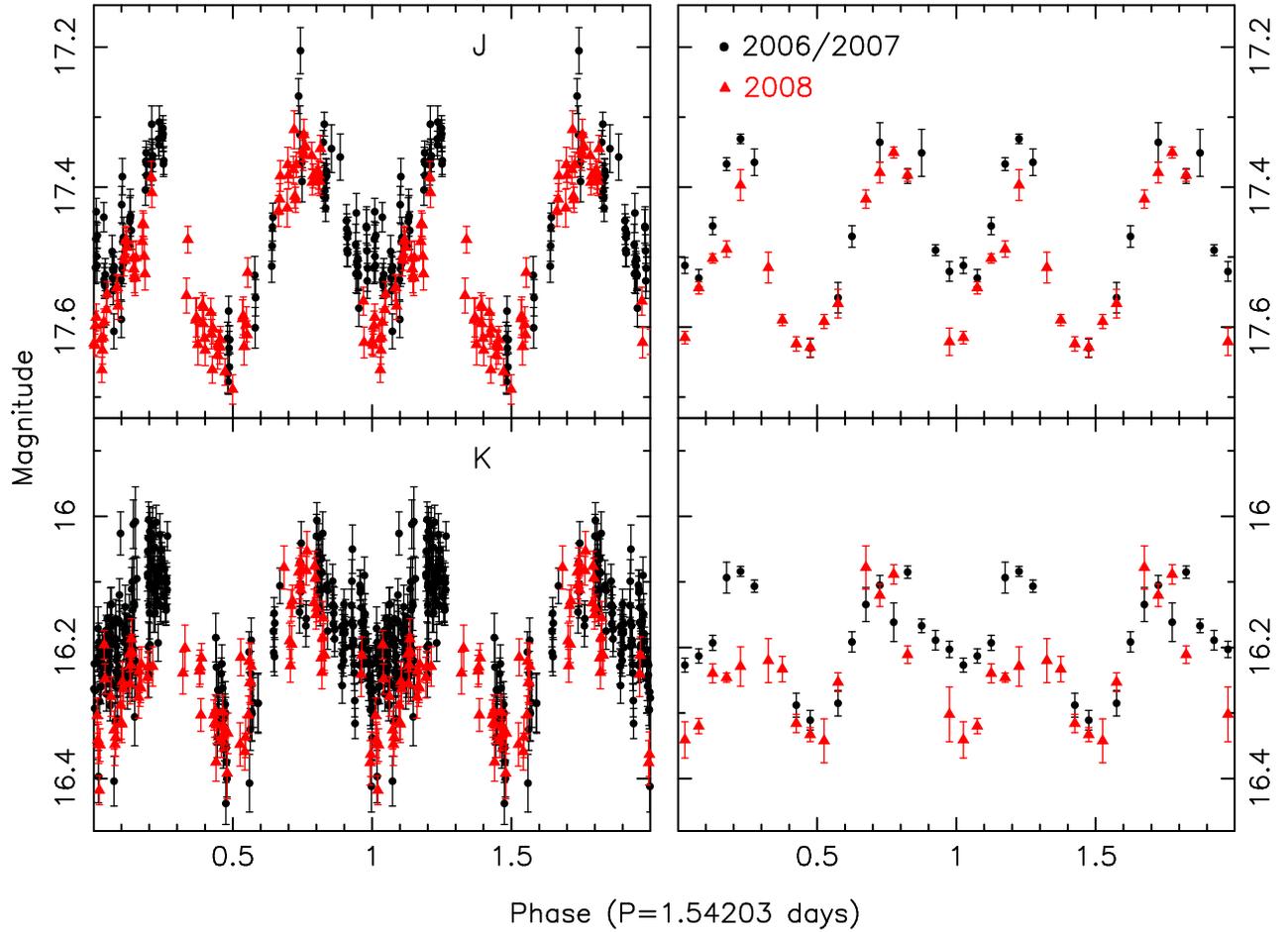}
\caption{The phased $J$ (top) and $K_S$ (bottom) light curves of \xte,
where phase zero corresponds to the time of the inferior conjunction
of the secondary star.  The panels on the left show the individual
measurements, and the panels on the right show the binned light
curves.}
\label{plotallpanic}
\end{figure}

\clearpage

\begin{figure}
\epsscale{0.9}
\plotone{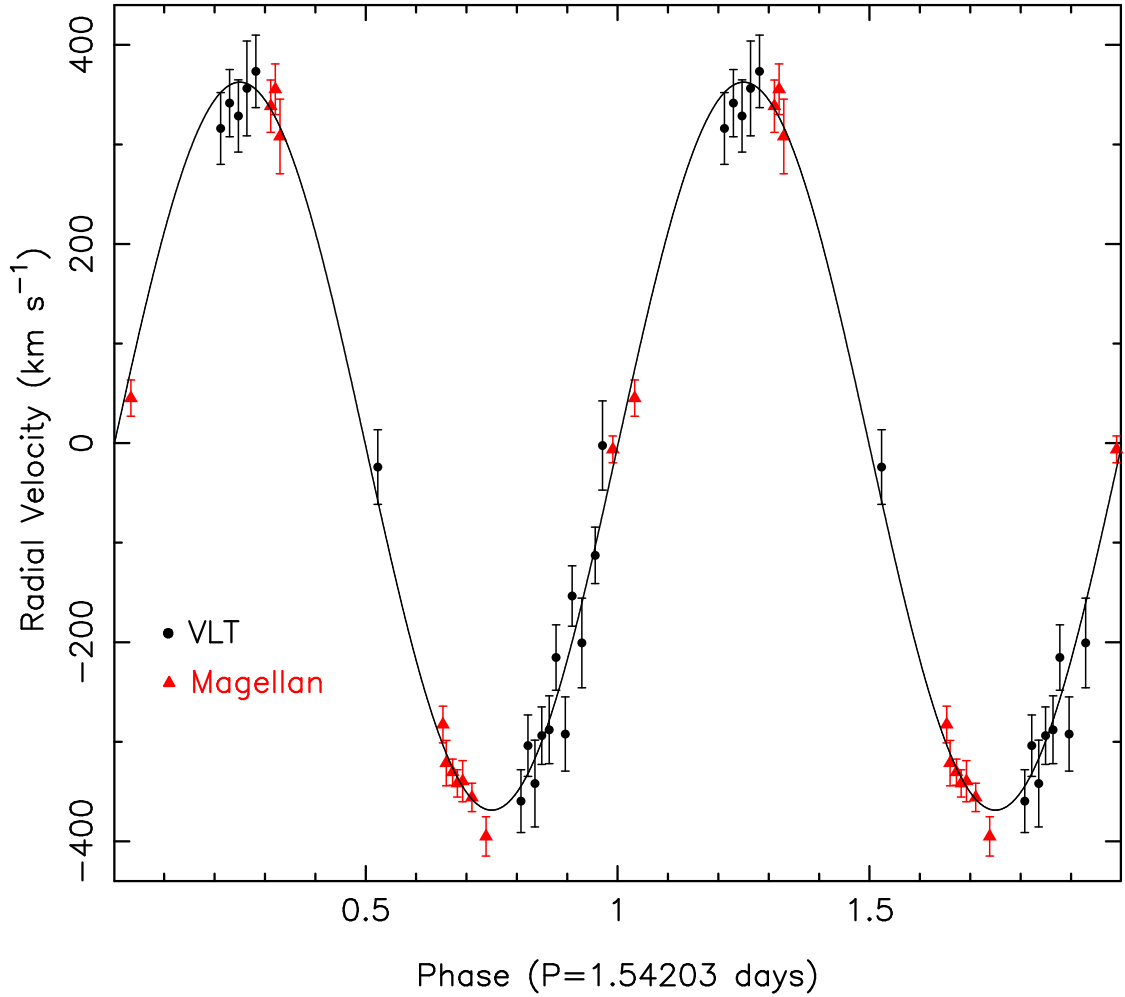}
\caption{The phased velocity curve of \xte, where phase zero
corresponds to the time of the inferior conjunction of the secondary
star.  The VLT radial velocities are shown with the filled circles
and the MagE radial velocities are shown with the filled triangles.
The solid line is the best-fit model assuming that the orbit is
circular.}
\label{rvfig1}
\end{figure}

\clearpage

\begin{figure}
\epsscale{0.9}
\plotone{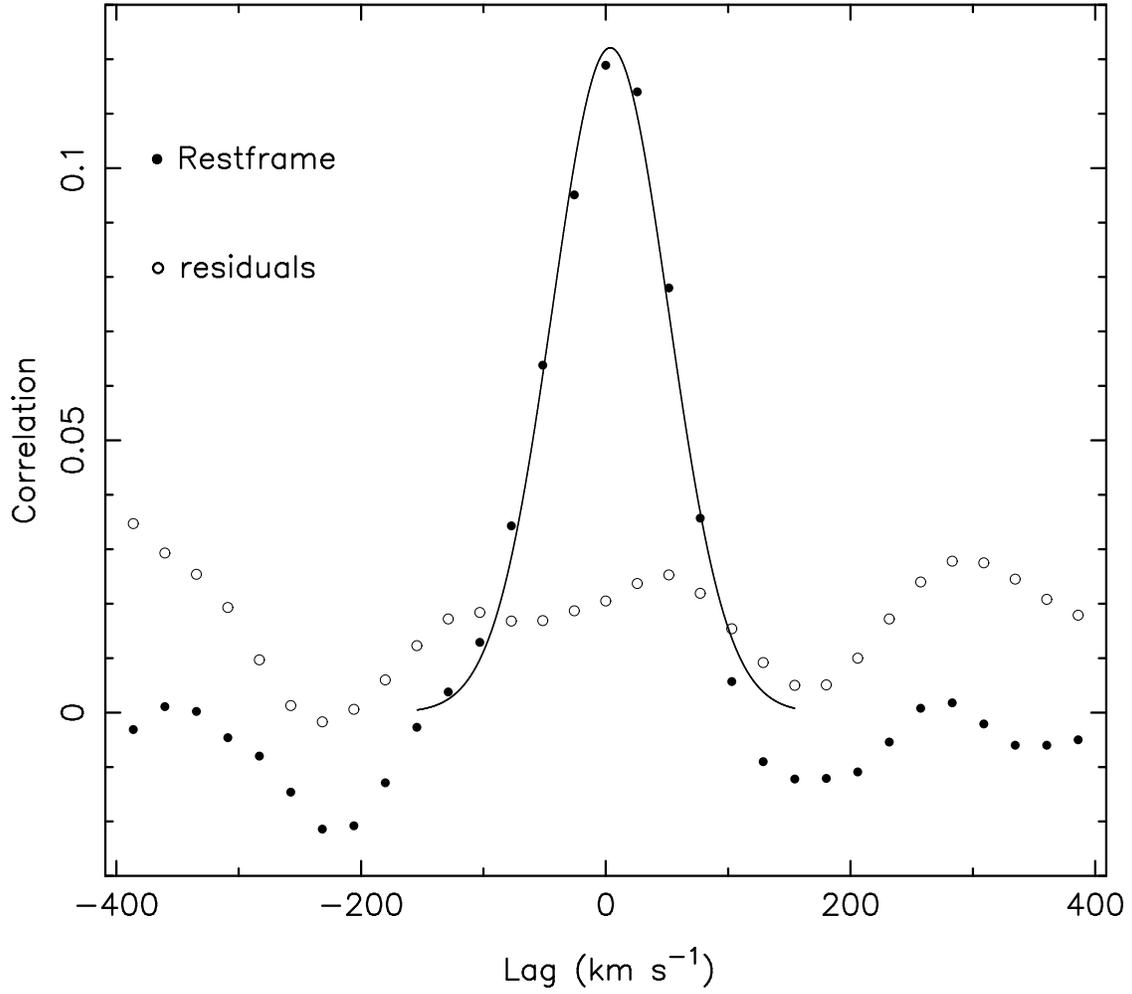}
\caption{The broadening kernel derived from the \xte\ restframe
spectrum is shown by the filled circles.  The best-fitting Gaussian
(with a FWHM of $65.4\pm 2.5$ km s$^{-1}$) is shown with a solid
line.  The broadening kernel derived from the residual spectrum
(i.e.\ the difference between the restframe spectrum and the scaled
spectrum of the template spectrum) is shown by the open circles.
There is no discernible signal in the kernel of the residual
spectrum, which is a good indication that the stellar absorption
lines have been mostly removed.
}
\label{BFfig}
\end{figure}

\clearpage

\begin{figure}
\epsscale{0.9}
\plotone{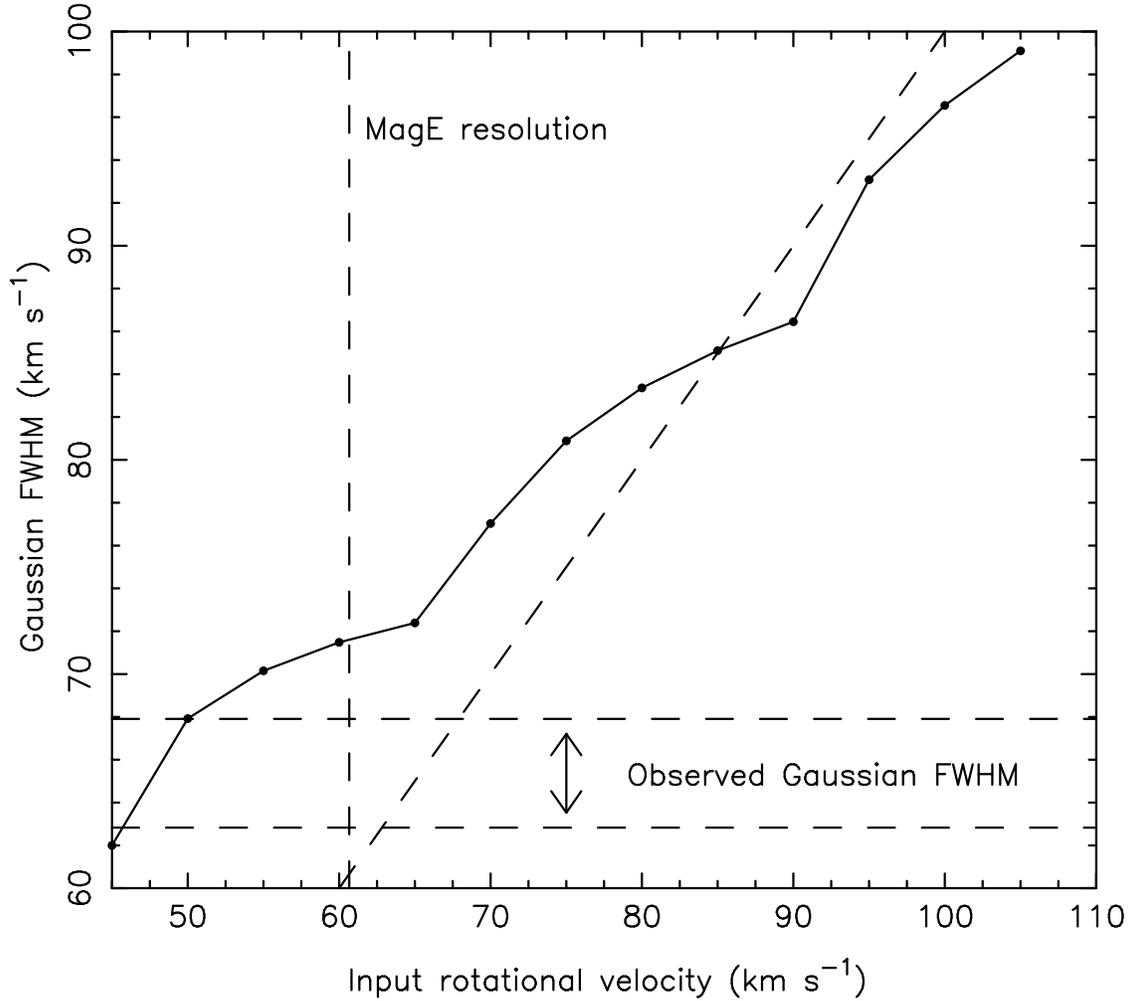}
\caption{The Gaussian FWHMs of broadened reference star kernels are
shown with filled circles.  The diagonal dashed line indicates where
the Gaussian FWHM equals the input rotational velocity.  The
vertical dashed line denotes the MagE spectral resolution, and the
horizontal dashed lines denote the observed $1\sigma$ range of the
Gaussian FWHM of the broadening kernel derived from the
\xte\ restframe spectrum (i.e.\ ${\rm FWHM}=65.4\pm 2.5$ km
s$^{-1}$).}
\label{calibfig}
\end{figure}

\clearpage

\begin{figure}
\epsscale{0.9}
\plotone{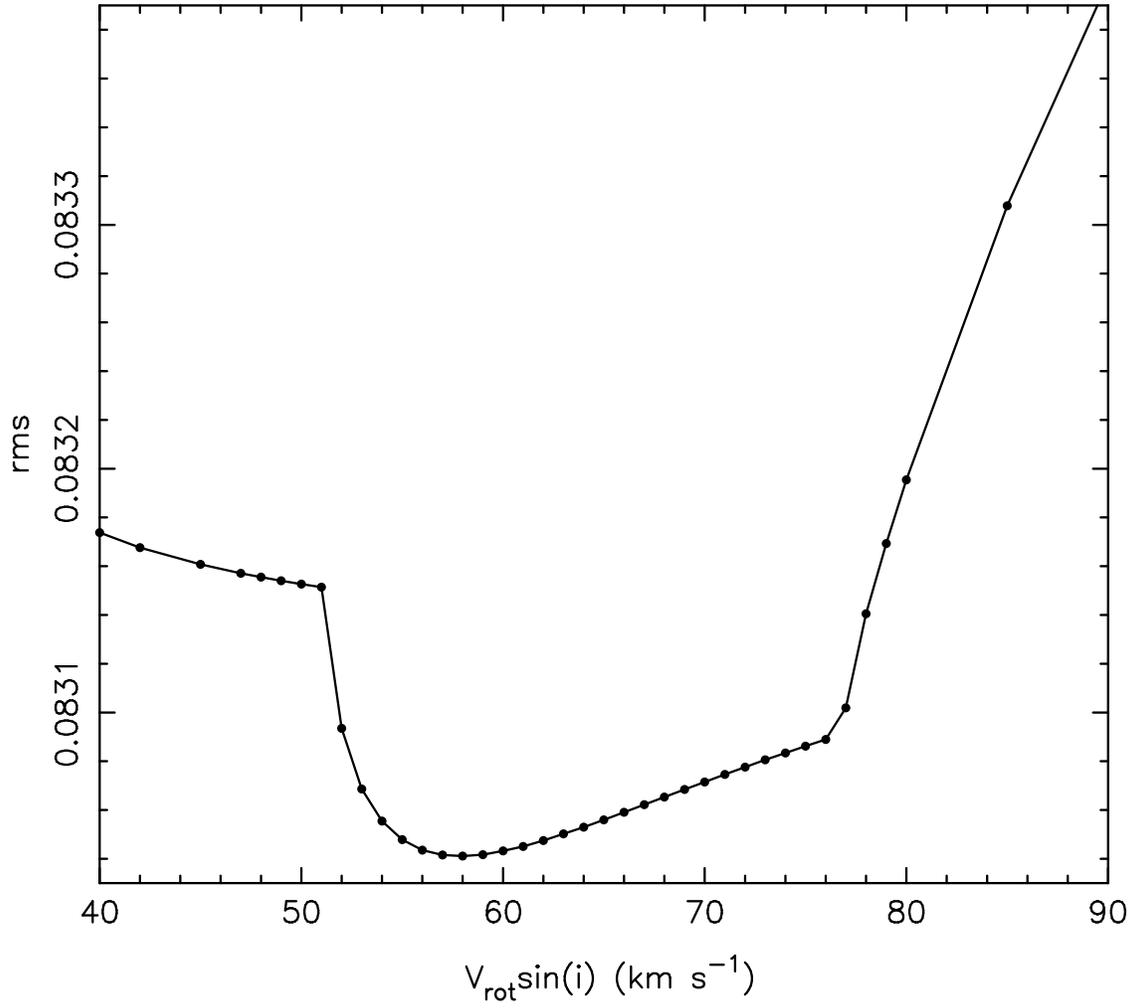}
\caption{The rms value of the polynomial fits to the difference
spectra as a function of the input projected rotational velocity of
the template spectrum (which is the K3III star HD 181110).  The
minimum rms occurs when $V_{\rm rot}\sin i=57$ km s$^{-1}$.}
\label{vrotfig}
\end{figure}

\clearpage

\begin{figure}
\epsscale{0.8}
\plotone{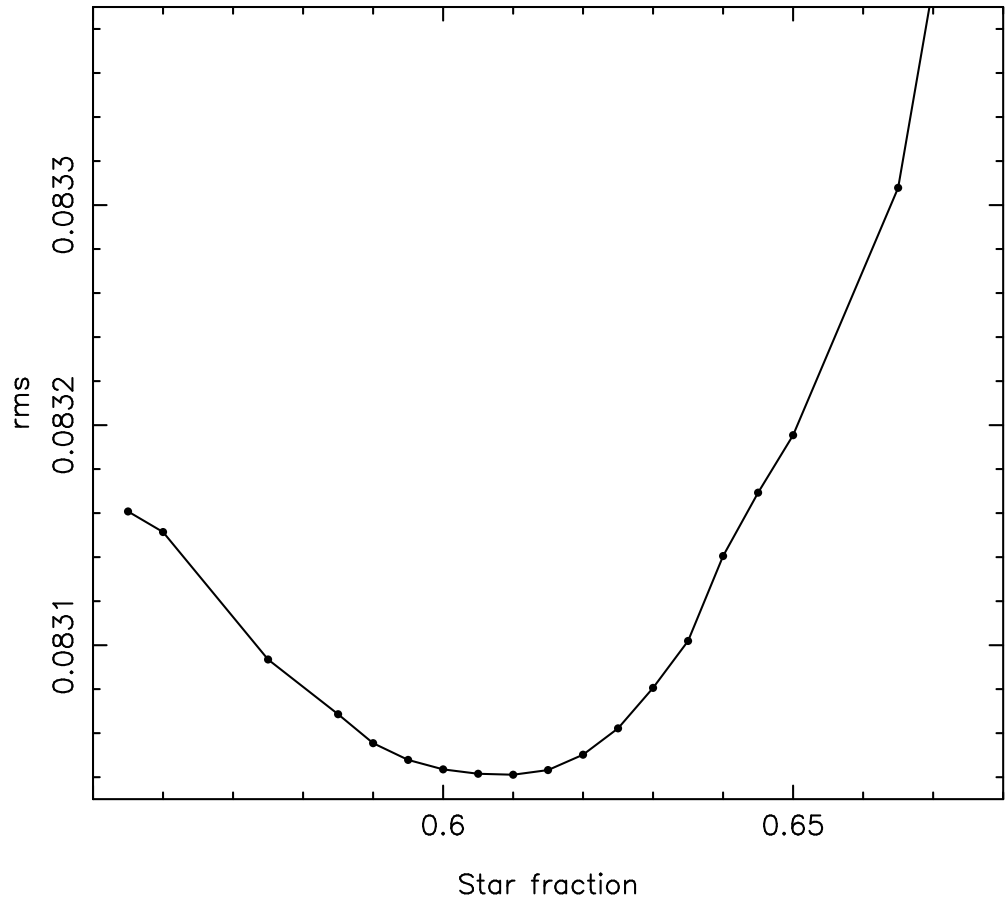}
\caption{The rms value of the polynomial fits to the difference
spectra as a function of the input star fraction, using the K3 III
star HD 181110 as the template spectrum.  We find a disk
fraction of 0.39 at an effective wavelength of 6200~\AA.}
\label{dffig}
\end{figure}

\clearpage

\begin{figure}
\epsscale{0.65}
\plotone{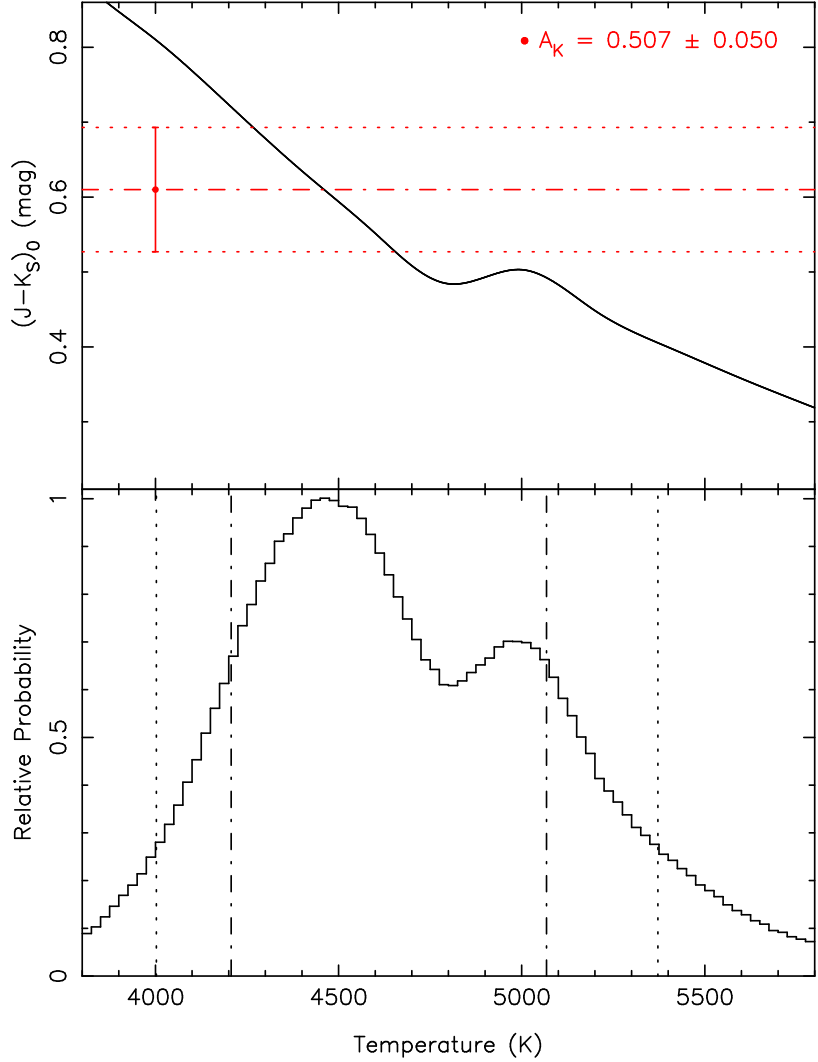}
\caption{Top: The solid line gives the unreddened $J-K_S$ color of the
secondary star as a function of temperature.  The measured value of
$(J-K_S)_0=0.610\pm 0.071$ assuming $A_K=0.507\pm 0.050$ is
indicated by the filled circle, and the $1\sigma$ range in the color
is shown by the red horizontal dashed lines.  Bottom: The
distribution of the allowed temperatures derived from the observed
color (assuming $A_K=0.507\pm 0.050$) is shown as the histogram.
The 68\% and 90\% confidence intervals are denoted by the vertical
dot-dashed lines and the dotted lines, respectively.  Considering
also the spectral type of the star, the most likely temperature is
about $T_{\rm eff}=4450$ K.}
\label{plottempvscolor}
\end{figure}

\clearpage

\begin{figure}
\epsscale{0.75}
\plotone{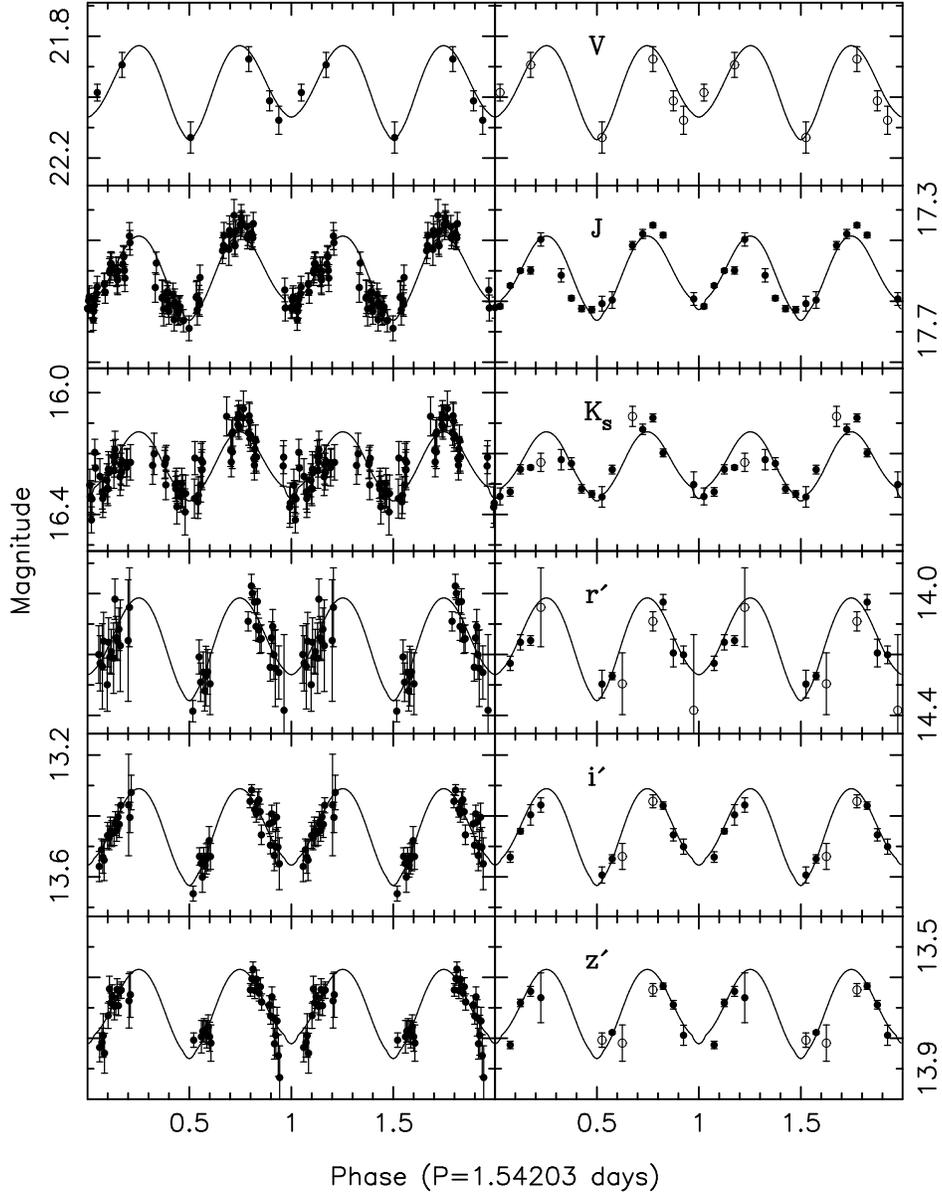}
\caption{The left panels show the folded light curves in the $V$, $J$,
$K_S$, and Sloan $r^{\prime}$, $i^{\prime}$, and $z^{\prime}$
filters and the best-fitting ellipsoidal models.  The smoothed light
curves shown in the right panels were made by computing the median
magnitude within bins 0.05 phase units wide.  Bins with a single
point are indicated by an open circle.  Note that only the $V$, $J$,
and $K_S$ magnitudes are calibrated to the standard scales.}
\label{fitfig}
\end{figure}

\clearpage

\begin{figure}
\epsscale{0.8}
\plotone{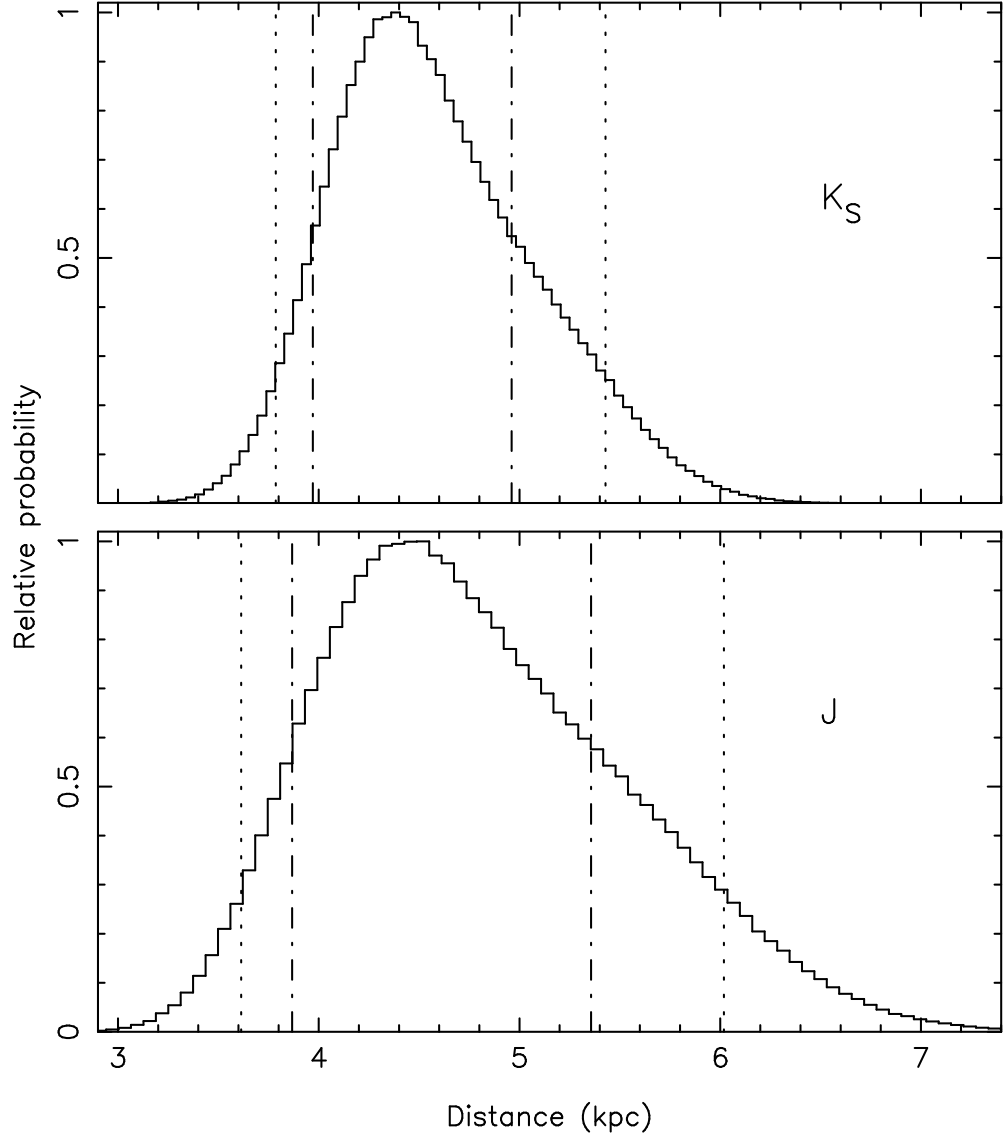}
\caption{Top: This plot shows the probability distribution of the
distance in kpc derived from a Monte Carlo simulation using the
apparent $K_S$ magnitude and extinction (see text).  The 68\% and
90\% confidence intervals are denoted by the vertical dot-dashed
lines and the dotted lines, respectively.  The most likely distance
is 4.38 kpc, with a $1\sigma$ range of $3.97 \le d \le 4.96$.
Bottom: Same as the top, but with the distance computed using
$J$-band measurements.  The most likely value is 4.52 kpc, with a
$1\sigma$ range of $3.87\le d\le 5.36$.}
\label{distfig}
\end{figure}

\end{document}